\documentclass{aa}
\usepackage[varg]{txfonts}
\usepackage{graphicx,lscape}
\bibpunct{(}{)}{;}{a}{}{,}

\newcommand{\msano}{{\rm M}_\odot ~{\rm yr}^{-1}}
\newcommand{\mdot}{\dot{M}}

\newcommand{\e}[1]{\times 10^{#1}}
\newcommand{\vtau}{V830\,Tau} 
\newcommand{\dd}{\rm d}

\begin{document}

\title{Predicting radio emission from the newborn hot Jupiter  \vtau\ b and its host star}
\titlerunning{Predicting radio emission from  \vtau\ b and its host star}
\authorrunning{Vidotto and Donati}
\author{A.~A.~Vidotto\inst{1}
\and J.-F.~Donati \inst{2,3}
}

\institute{School of Physics, Trinity College Dublin, University of Dublin, Ireland \and
Universit\'e de Toulouse, UPS-OMP, IRAP, 14 avenue E. Belin, Toulouse F-31400, France 
\and 
CNRS, IRAP / UMR 5277, 14 avenue E. Belin, Toulouse F-31400, France}


\date{Received date /
Accepted date }

\abstract{
Magnetised exoplanets are expected to emit at radio frequencies analogously to the radio auroral emission of Earth and Jupiter. Here, we predict the radio emission from \vtau\ b, the youngest (2 Myr) detected exoplanet to date. We model the wind of its host star using three-dimensional magnetohydrodynamics simulations that take into account the reconstructed stellar surface magnetic field. Our simulations allow us to constrain the local conditions of the environment surrounding \vtau\ b that we use to then compute its radio emission. We estimate average radio flux densities of 6 to 24~mJy, depending on the assumption of the radius of the planet (one or two Jupiter radii). These radio fluxes are not constant along one planetary orbit, and present peaks that are up to twice the average values. We show here that these fluxes are weakly dependent (a factor of 1.8) on the assumed polar planetary magnetic field (10 to 100~G), opposed to the maximum frequency of the emission, which ranges from 18 to 240 MHz. We also estimate the thermal radio emission from the stellar wind. By comparing our results with the Karl G. Jansky Very Large Array and the Very Long Baseline Array observations of the system, we constrain the stellar mass-loss rate to be $\lesssim 3\e{-9}~\msano$, with likely values between $\sim 10^{-12}$ and $10^{-10}~\msano$. With these values, we estimate that the frequency-dependent extension of the radio-emitting wind is around $\sim 3$ to 30 stellar radii ($R_\star$) for frequencies in the range of 275 to 50 MHz, implying that \vtau\ b, at an orbital distance of $6.1~R_\star$, could be embedded in the regions of the host star's wind that are optically thick to radio wavelengths, but not deeply so.  We also note that planetary emission can only propagate in the stellar wind plasma if the frequency of the cyclotron emission exceeds the stellar wind plasma frequency. In other words, we find that for planetary radio emission to propagate through the host star wind, planetary magnetic field strengths larger than $\sim 1.3$ to $13$ G are required. Since our radio emission computations are based on analogies with solar system planets, we caution that our computations should be considered as estimates. Nevertheless, the \vtau\ system is a very interesting system for conducting radio observations from both the perspective of radio emission from the planet as well as from the host star's wind.
}

\maketitle

\section{Introduction}\label{sec.intro}
Detecting radio emission from exoplanets would revolutionise the area of exoplanetary studies. First, it would open up a new avenue for the direct detection of exoplanets \citep{1999JGR...10414025F}. Second, it would allow us to measure exoplanetary magnetic fields, which so far have only been elusively probed \citep{2008ApJ...676..628S,2010ApJ...722L.168V,2014Sci...346..981K}. Exoplanetary magnetic fields can reveal information about a planet's interior structure and dynamics (e.g. whether it has a convecting, electrically conducting interior) and are also believed to play an important role in conditions for habitability, by protecting the planet's surface from energetic cosmic particles, protecting its atmosphere from violent chemical changes, and potentially helping atmospheric retention \citep{2007SSRv..129..245L,2016A&A...587A.159G}.

The magnetised planets in the solar system generate radio emission through the electron cyclotron maser instability. The power of this emission is related to the kinetic and/or magnetic powers of the incident solar wind over several orders of magnitude \citep{1999JGR...10414025F,2001Ap&SS.277..293Z}. Although the physics of this relation are still not well understood, this ``radiometric Bode's relation'' indicates that planetary radio emission is somehow powered by the interaction between the planetary magnetic field and the solar wind. In analogy to the solar system, it has been suggested that exoplanets may also produce radio emission due to their interaction with the winds of their host stars \citep{1999JGR...10414025F}.\footnote{In particular, \citet{2010ASPC..430..175Z,2015aska.confE.120Z} found that the scaling between the magnetic power of the incident stellar wind on the  magnetospheric cross-section and the radio power apply also to any plasma flow-obstacle interaction, including interactions between Io and Ganymede with Jupiter and the magnetised binary system V711 Tau. This generalised radio-magnetic Bode's law covers about 13 orders of magnitude in stellar wind magnetic power and radio power. } 

Calculations based on the solar-system analogy suggest that radio emission from close-in exoplanets should be  several  orders of magnitude larger than the largest radio emitter of the solar system, Jupiter \citep[e.g.,][]{2001Ap&SS.277..293Z,2004ApJ...612..511L,2005A&A...437..717G,2007P&SS...55..598Z,2010ApJ...720.1262V,2010MNRAS.406..409F, 2011A&A...531A..29H}. This is because, at the short orbital distances of close-in planets, the local power of  the host star wind is significantly higher than, for example, the local power of the solar wind dissipated at the distances of the planets in our solar system. These and many other theoretical predictions have motivated a good number of observational searches of exoplanetary radio emission, yielding mostly negative results (e.g. \citealt{2000ApJ...545.1058B,2004P&SS...52.1479R, 2007ApJ...668.1182L,2013ApJ...762...34H}, but see also \citealt{2013A&A...552A..65L,2014A&A...562A.108S}, who found ambiguous hints of the existence of exoplanetary radio emissions). There are several reasons for the non-detections, such as a lack of exoplanetary magnetism, a mismatch of the frequency search, and/or a low sensitivity of the present-day instruments (\citealt{2000ApJ...545.1058B}, \citealt{2011RaSc...46.0F09G}).

One might overcome the issue of instrumental sensitivity by turning to exoplanets that, not only orbit closer to their stars, but that orbit stars that have winds that are more powerful than solar-type winds, and whose magnetic fields are also stronger. Stars that are more active than the Sun are indeed expected to host more powerful winds \citep{2004LRSP....1....2W,2014A&A...570A..99S,2015A&A...577A..28J,2016ApJ...820L..15D}. In particular, stars in the pre-main sequence with close-in planets would be ideal targets to search for planetary radio emission \citep{2005A&A...437..717G, 2010ApJ...720.1262V}. With the recent detection of the first hot giants orbiting stars younger than $10$--$20$~Myr (\vtau\ b, \citealt{2016Natur.534..662D}, and K2-33b, \citealt{2016Natur.534..658D}, TAP26b, \citealt{2017MNRAS.tmp...26Y}), now is an excellent timing for testing theoretical expectations of exoplanetary radio emission. In this paper, we compute the expected radio emission from the newborn hot Jupiter \vtau\ b. 

With the spectropolarimetric monitoring of \vtau , \citet{2016Natur.534..662D} were able to extract the planetary signature of \vtau\ b, a 2~Myr-old hot Jupiter orbiting at 0.057~au, while, at the same time, reconstructing the large-scale magnetic field topology of the host star. It is precisely the latter measurements that we use in this paper to derive the wind properties of the host star (Section~\ref{sec.windmodel}). With that, we are then able to derive the expected radio emission of \vtau\ b, starting from the assumption that \vtau\ b is magnetised (Section~\ref{sec.radiomodel}).  Given the high densities of the wind of young active stars, these winds can become optically thick to free-free radiation at radio wavelengths. We thus investigate in which circumstances the planet would be embedded in the radio-emitting region of the stellar wind and  whether planetary radio emission would be able to propagate through the stellar wind plasma, given that the planetary cyclotron frequency of emission should exceed the plasma frequency of the stellar wind (Section \ref{sec.radiowind}). For physically-reasonable conditions, we show that planetary radio emission from \vtau\ b is expected to peak at 12~mJy, presenting therefore a high potential of detection with LOFAR (the Low-Frequency Array) the upgraded UTR-2 (Ukrainian T-shaped Radio telescope) and GMRT (Giant Metrewave Radio Telescope ,see Section~\ref{sec.conclusion}). Interestingly, its host star has been detected using VLA (the Karl G. Jansky Very Large Array) at 4.5 and 7.5~GHz and with VLBA (the Very Long Baseline Array) at 8.4~GHz, becoming the first exoplanet host with detected radio emission \citep{bower}.

\section{Stellar wind modelling}\label{sec.windmodel}
We model the wind of \vtau\ by means of three-dimensional (3D) magnetohydrodynamics (MHD) simulations, similarly to \citet{2012MNRAS.423.3285V,2015MNRAS.449.4117V}. We refer the reader to these papers for more details of the model, which are summarised next. We use the 3D MHD numerical code BATS-R-US \citep{1999JCoPh.154..284P,2012JCoPh.231..870T}, modified as in \citet{2012MNRAS.423.3285V}. BATS-R-US solves the set of ideal MHD equations for the mass density,  velocity, magnetic field, and  gas pressure. We assume the wind is polytropic with a polytropic index $\gamma=1.15$ and  a fully ionised hydrogen wind. 

For the physical characteristics of the star, we use a rotation period of $2.741$~days, mass of $1 M_\odot$,  and radius of $R_\star=2 R_\odot$. The radial part of the stellar magnetic field, anchored at the wind base, is constrained from observations collected in November-December 2015 \citep{2016Natur.534..662D}. Initially, the field is considered to be in its lowest (potential) state, but  as the simulation evolves in time the wind particles self-consistently interact with the magnetic field lines (and vice-versa), removing the field from its initial potential state. At the wind base, we adopt a temperature of $10^6$~K and a number density of $n_0=10^{12}$cm$^{-3}$, which are free parameters of our model. The star rotates as a solid body with a rotation axis along the $z$-axis. Our grid is the same as described in \citet{2014MNRAS.438.1162V}. The resultant wind solution, obtained self-consistently, is found when the system reaches steady state in the reference frame corotating with the star. For the parameters we adopted, we obtain a wind mass-loss rate of $\mdot \sim 3\e{-9}~\msano$. This mass-loss rate is an upper limit for the case \vtau , as we will demonstrate in Section \ref{sec.radiowind}.

\begin{figure}[ht!]
        \includegraphics[width=\columnwidth]{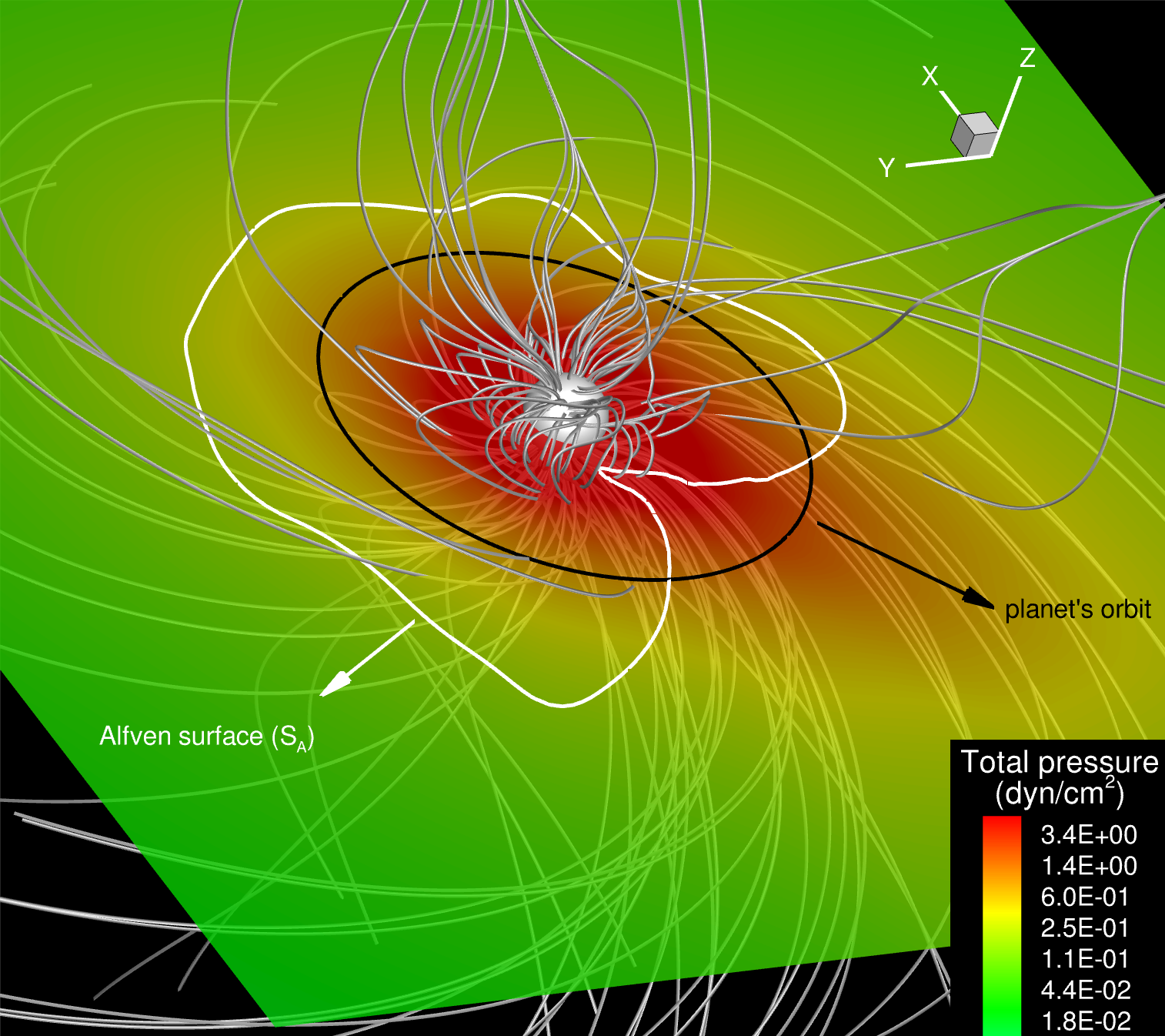}%
\caption{Simulated wind of \vtau . Thin grey lines represent the magnetic field of \vtau\ that is embedded in the wind. The circle depicts the orbital radius of \vtau\ b, assumed to lie in the equatorial plane of the star ($xy$ plane). The white line represents a cut of the Alfv\'en surface at the equatorial plane, while the colour shows the total wind pressure.}\label{fig.wind}
\end{figure}

Figure \ref{fig.wind} illustrates the output of the wind simulations of \vtau . We overplot the (assumed equatorial and circular) orbital radius of \vtau\ b (black circle) and  a cut of the Alfv\'en surface $S_A$ at the equatorial plane ($xy$ plane). For the parameters we adopted, the planet's orbit lies almost completely inside $S_A$. The Alfv\'en surface is expected to expand for lower values of $n_0$ \citep{2009ApJ...699..441V}. This indicates that \vtau\ b is truly  orbiting in the sub-Alfv\'enic regime, if not at all times, at least during most of its orbit. The Alfv\'en surface represents the boundary between a magnetically-dominated stellar wind (interior to $S_A$) and a region that is dominated by the wind inertia. The Alfv\'en surface is relevant for computing angular momentum losses \citep{1967ApJ...148..217W} and indicates the types of star-planet interactions. Planets orbiting in the sub-Alfv\'enic regime can be directly connected to the magnetic field of the star and perturbations can travel to/from the star \citep[e.g.][]{2005A&A...434.1191P,2013A&A...552A.119S,2015ApJ...815..111S}. In this case, the planetary magnetosphere is embedded in that of the star and it is believed that there is continuous reconnection of the planet's and stellar magnetic fields, similar to the Ganymede-Jupiter system.
 When the planet orbits outside $S_A$ (in the super-Alfv\'enic regime), as in the case of the Earth, the magnetospheric cavity surrounding the planet deflects the stellar wind particles. We estimate the size of the magnetosphere of \vtau\ b in the next section.  
 
\section{Exoplanetary radio emission} \label{sec.radiomodel}
To calculate the radio emission of \vtau\ b, we use the radiometric Bode's law for the solar system, in which the planetary radio power $P_{\rm radio}$  can be decomposed into a power released from the dissipation of kinetic energy of the stellar wind $P_k$ and/or a power released from the dissipation of magnetic energy of the wind $P_B$ \citep[e.g.][]{1999JGR...10414025F,2001Ap&SS.277..293Z}
\begin{equation}\label{eq.pwrrec}
P_{\rm radio} = \eta_k P_k  \,\,\,\,  \,\,\,\,  \,\,\,\, \textrm{or }  \,\,\,\,  \,\,\,\,  \,\,\,\, P_{\rm radio} =  \eta_B P_B\, ,
\end{equation}
where $\eta_k $ and $\eta_B$ are efficiency ratios. In the solar system, $\eta_k  = 1\times 10^{-5}$ and $\eta_B=2\times 10^{-3}$ \citep{2007P&SS...55..598Z}; we assume the same efficiency ratios here. In the past, it has been argued that it was not possible to decide which incident power actually drives the radio power observed from the magnetic planets of the solar system \citep{2007P&SS...55..598Z}. Recently, however, \citet{2010ASPC..430..175Z} argued that the magnetic field is likely to be a determinant for extracting part of the flow power and converting it to energetic particles \citep[see e.g.][for a theoretical basis of this empirical relation]{2008A&A...490..843J,2016MNRAS.461.2353N}. Here, for completeness, we compute the radio powers coming from both kinetic and magnetic energies. As we will see, the latter is much larger than the former and, following the arguments in \citet{2010ASPC..430..175Z}, is the preferred method for estimating radio fluxes. 

\begin{figure}
        \includegraphics[width=0.98\columnwidth]{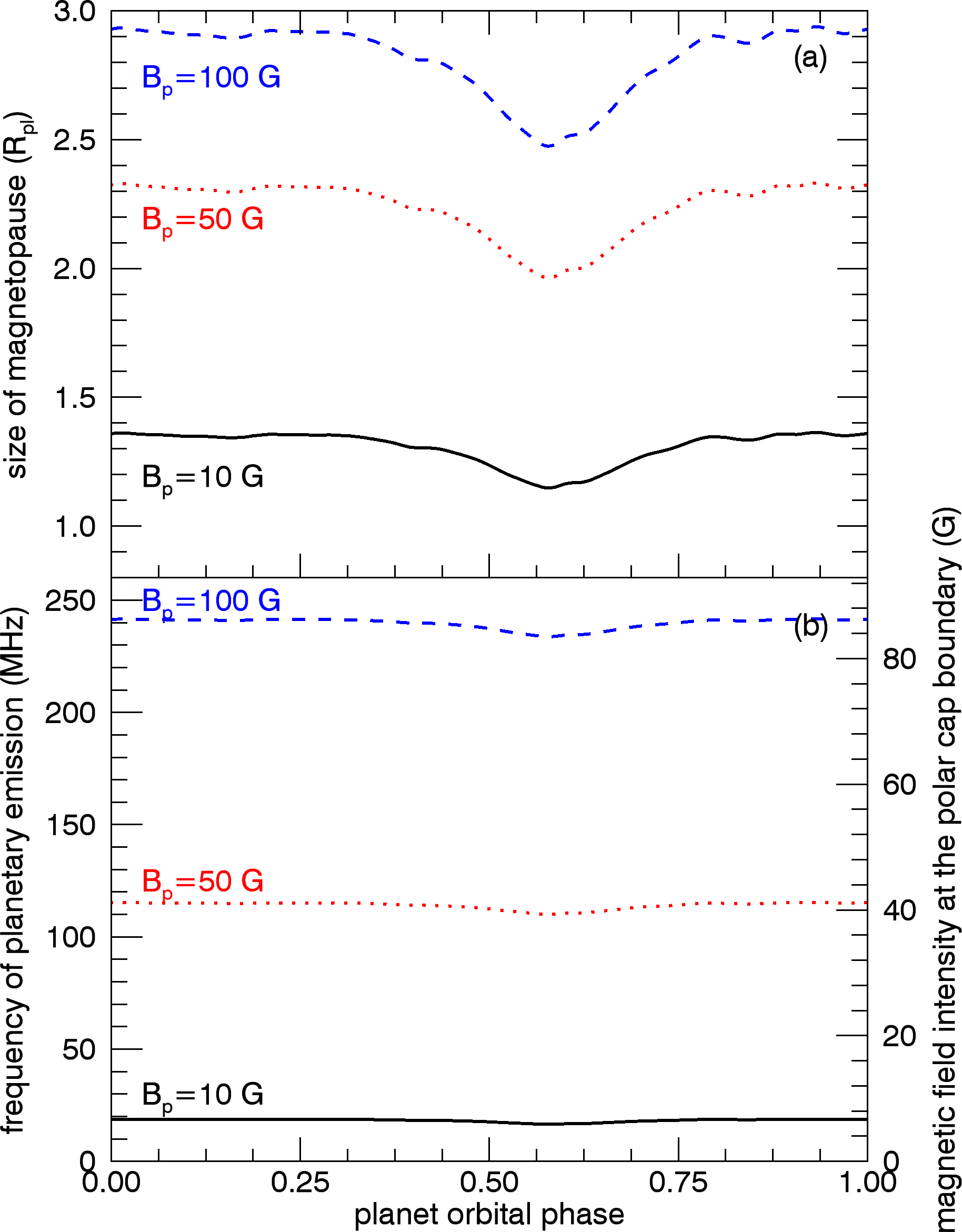}%
\caption{(a) Estimated sizes of the planetary magnetospheres along a planetary year assuming three different values of the polar planetary magnetic field: $10$, $50,$ and $100$~G. (b) Predicted maximum frequency of the radio emission of \vtau\ b. This frequency is calculated at the border of the polar-cap boundary, which is located at different colatitudes depending on the size of the planet magnetosphere (\textit{cf}, Eq.~\ref{eq.alpha0}). The magnetic field intensity at this position in shown in the right axis.}\label{fig.rm}
\end{figure}

The dissipated kinetic and magnetic powers of the impacting wind on the planet are approximated as, respectively,  (Appendix \ref{sec.apA})
\begin{eqnarray}
P_k \simeq \rho (\Delta u)^3 \pi r_M^2  \, , && \label{eq.pK} \\
P_B  \simeq  \frac{B_{\perp}^2 }{4\pi} (\Delta u) \pi r_M^2\, , && \label{eq.pB}
\end{eqnarray}
where $B_{\perp}$ is the magnetic field component perpendicular to $\Delta u$. Here,  $|\Delta {\bf u}| = |{\bf u} - {\bf v}_{K}|$ is the relative velocity between the wind and the Keplerian velocity ($v_K$) of the planet, and $\rho$, $p$ , and $B$ are the local density, pressure, and magnetic field intensity of the stellar wind. Neglecting planet thermal pressure, the size of the planet's magnetopause $r_M$ can be estimated by  pressure balance between the stellar wind total pressure and the planet's magnetic pressure: $p_{\rm tot}= {B_{p}^2(r_M)}/({8\pi})$, where $p_{\rm tot} = {\rho \Delta u^2} + \frac{B^2}{8\pi} + p$ is the sum of the ram, magnetic, and thermal pressures of the wind;  $B_{p}(r_M)$ is the intensity of the planet's magnetic field at the nose of the magnetopause. For a dipolar planetary magnetic field: $B_{p}(r_M) = \frac12 B_p (R_p/R)^3$, where $R_p$ is the planetary radius, $R$ is the radial coordinate centred at the planet, and $B_{p}$ is the polar magnetic field intensity (at the equator, the intensity is $\frac12  B_p $). Assuming the dipole is aligned with the planetary orbital spin axis, the magnetospheric radius (where $R=r_M$) is
\begin{equation}\label{eq.rM}
\frac{r_M}{R_p}= 2^{1/3}\left[ \frac{(B_p/2)^2/{8\pi} }{{\rho \Delta u^2}  + p + B^2/{8\pi} } \right]^{1/6} .
\end{equation}
Figure \ref{fig.rm}a shows the stand-off distance of \vtau\ b's magnetopause calculated using Eq.~(\ref{eq.rM})  as the planet orbits around the star.\footnote{The right hand side of Eq.~(\ref{eq.rM}) is often multiplied by a correction factor  $2^{1/3}$ used to account for the effects of currents \citep[e.g.][]{2004pssp.book.....C}.} We assume three different values of $B_{p}$: $10$, $50,$ and 100~G. For comparison, we note that the maximum intensity of Jupiter's magnetic field  is $14.3$~G \citep{1992AREPS..20..289B}. For $B_{p} = 10$ G, the average  magnetospheric radius is $\sim 1.31R_p$ (i.e. very close to the planet surface), while in the second and third cases, $r_M \sim 2.2R_p$ and $2.8 R_p$, respectively. If the wind densities (and therefore mass-loss rates) were to decrease by a factor of $100$, the magnetospheric sizes would increase by about $20\%$ of the values presented in Figure \ref{fig.rm}a (in such a case, the planet's orbit would lie completely within the Alfv\'en surface). This shows that, even in the case of a  ``high'' planetary magnetic field and low wind densities, the magnetosphere of \vtau\ b is expected to be quite small, due to the harsh conditions of the external environment.
 

The radio flux is related to the radio power as
\begin{equation}
\phi_{\rm radio} = \frac{P_{\rm radio}}{d^2 \omega \Delta f} \, ,
\end{equation}
where $\omega$ is the solid angle of the (hollow) cone of emission,  $d$ is the distance to the system ($147$~pc in the case of \vtau ), and $\Delta f$ is the emission bandwidth, approximately the cyclotron frequency  \citep{2007P&SS...55..618G}. In Appendix \ref{sec.apA}, we show that the radio flux density due to the dissipated wind kinetic and magnetic powers simplify to
\begin{eqnarray}
\phi_{\rm radio, kin}  = \eta_k^\prime \frac{R_p^2}{d^2 }\frac{\rho (\Delta u)^3}{{{ p_{\rm tot}}^{1/2}}}  f(\alpha_0) , \label{eq.phiradio_maintext} \\
\phi_{\rm radio, mag}  = \eta_B^\prime \frac{R_p^2}{d^2 }\frac{B_\perp^2 (\Delta u)}{{{ p_{\rm tot}}^{1/2}}}  f(\alpha_0) , \label{eq.phiradio_B_2_maintext}
\end{eqnarray}
respectively. Here $\eta_k^\prime$  and $\eta_B^\prime$ are constants and are related to $\eta_k$ and $\eta_B$ following Eqs.~(\ref{eq.etakprime}) and (\ref{eq.etaBprime}). The planetary magnetic field dependence is hidden in the function $f(\alpha_0)$, which varies between $0$ and $3.3$ for any planetary magnetic field intensity (Figure~\ref{fig.function_variation}). In particular, we assume that the beaming angle of the radio emission occurs at a ring with colatitude $\alpha_0$ and thickness $\delta\alpha = 17.5^{\rm o}$ \citep{2004JGRA..10909S15Z}.  Using the values for $\eta_k$ and $\eta_B$ from the solar system, we have $ \eta_k^\prime \simeq  1.8 \e{-13} \textrm{\, [cgs~units]}$ and $\eta_B^\prime \simeq 2.8\e{-12} \textrm{\, [cgs~units]}$. From Eqs.~(\ref{eq.phiradio_maintext}), (\ref{eq.phiradio_B_2_maintext}), (\ref{eq.etakprime}), and (\ref{eq.etaBprime}), we have that
\begin{equation}\label{eq.ratio}
\frac{\phi_{\rm radio, kin} }{\phi_{\rm radio, mag} } = \frac{\eta_k}{\eta_B (4\pi)^2} \frac{\rho (\Delta u)^2/2}{B_\perp^2/8\pi} = \frac{1}{3200 \pi^2} \frac{\rho (\Delta u)^2/2}{B_\perp^2/8\pi} .
\end{equation}
Equation~(\ref{eq.ratio}) shows that the flux released in the dissipation of Poynting flux becomes dominant when the kinetic energy ($\rho (\Delta u)^2/2$) of the impacting wind is smaller than $3200 \pi^2$ times the magnetic energy ($B_\perp^2/8\pi$). This is the case of the \vtau\ system, due to the intense stellar magnetism.

From Eqs.~(\ref{eq.phiradio_maintext}) and (\ref{eq.phiradio_B_2_maintext}), we see that the radio fluxes are functions of the angular size of the planet ($R_p/d$) and the properties of the stellar wind surrounding the planet. We also note that radio fluxes depend relatively weakly on the magnetic intensity of the planetary dipolar field. 
This dependence is hidden in $f(\alpha_0)$ (\textit{cf} Appendix~\ref{sec.apA}). For example, for polar magnetic fields of 10, 50, and 100~G, respectively, the (northern) polar-cap boundaries are located at colatitudes $\alpha_0$ of $61^{\rm o}$, $42^{\rm o}$ , and $37^{\rm o}$, respectively. For these values of  $\alpha_0$, $f(\alpha_0)$ is $2.2$, $1.4,$ and $1.2$, respectively. Therefore, as radio fluxes are proportional to $f(\alpha_0)$, a change of field intensities from 10 to 100~G results in a rather small change (a factor of 1.8) in the computed radio fluxes. This is  good news, given that exoplanetary magnetism remains observationally elusive (\textit{cf} Sect. \ref{sec.intro}).

\begin{figure}
        \includegraphics[width=0.98\columnwidth]{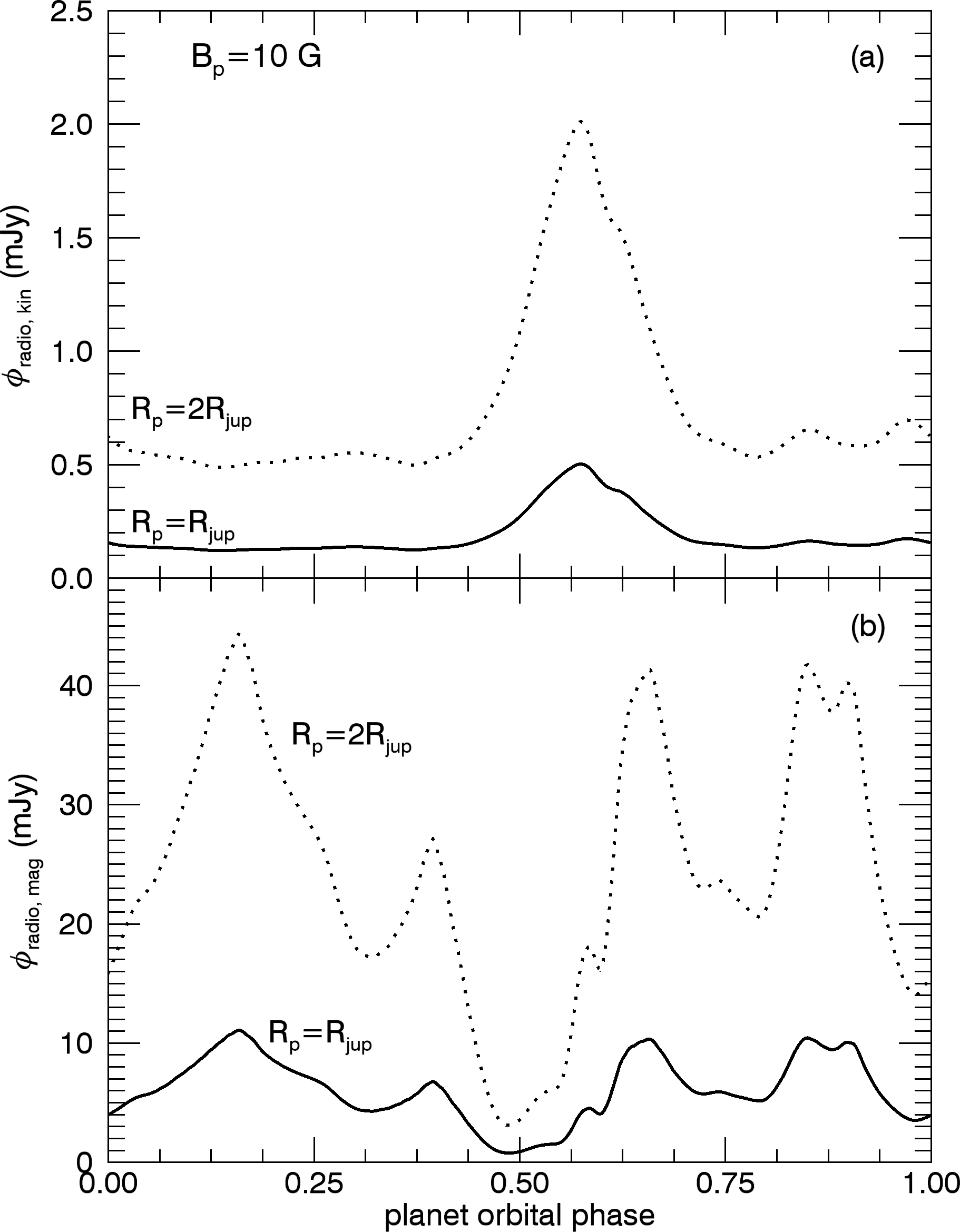}%
\caption{Predicted radio flux density along a planetary year assuming dissipation of kinetic (top) and magnetic (bottom) stellar wind powers onto the magnetospheric cross section of the exoplanet. At the low frequency range, values of a few mJy are potentially observable with LOFAR,  the upgraded UTR-2 and GMRT.}\label{fig.radio}
\end{figure}

Figure \ref{fig.radio} shows the predicted radio flux densities along a planetary year assuming dissipation of kinetic (top) and magnetic (bottom) stellar wind powers.  We found that radio emission arising from the dissipation of wind magnetic power is considerably larger  ($ \sim 30$ times) than from kinetic power and this difference could become even larger should a lower stellar mass-loss rate be adopted.  This is due to the kinetic-to-magnetic ratio being smaller than $3200\pi^2$ (\textit{cf} Eq.~\ref{eq.ratio}).
In Appendix \ref{sec.apB}, we show that our radio predictions for \vtau\ b are robust against a reduction of stellar wind densities and mass-loss rates, given the high magnetism of \vtau . We assume here that the planet has a radius of $1$ to $2~R_{\rm jup}$, but we note that, although the mass of \vtau\ b has been constrained by observations to be about $0.7~M_{\rm jup}$ \citep{2016Natur.534..662D}, no constraints on its radius currently exist. Since radio fluxes are proportional to $R_p^2$, the fluxes calculated for $R_p=2R_{\rm jup}$ are four times larger than for the calculations with $R_p=R_{\rm jup}$. For the case of $R_p=R_{\rm jup}$, the estimated magnetic radio fluxes are on average $\sim 6$~mJy, with peaks at $11$~mJy. This can be considered as a conservative estimate. 
Due to its youth and the close distance to the host star, it is likely that the radius of \vtau\ b is larger than $1R_{\rm jup}$, since \vtau\ b is likely to be still in the contracting phase \citep[e.g. K2-29 b, which has a similar mass, but orbits an older, 450~Myr-old star, has a radius of $1.19~R_{\rm jup}$, ][]{2016ApJ...824...55S}. For the case of $R_p=2R_{\rm jup}$, the estimated magnetic radio fluxes are on average $\sim 24$~mJy, with peaks at $44$~mJy.


We adopted  in Figure \ref{fig.radio}  a dipolar magnetic field whose polar intensity is $10$~G, but again note that an increase in magnetic field intensity by a factor of 10 would result in a change on the predicted radio flux by a factor of 1.8. The  planetary magnetic field has, however, a strong influence on the  frequency of emission. Since this is cyclotron emission, by measuring its frequency one is able to derive the intensity of the planetary magnetic field (\textit{cf} Eq.~\ref{eq.fcyc}, see \citealt{2011A&A...531A..29H} for a suggestion on how to conduct this estimation). Figure \ref{fig.rm}b shows the predicted maximum emission frequency, considering the same polar planetary magnetic fields as in Figure \ref{fig.rm}a. This frequency is calculated at the border of the polar-cap boundary, which is located at different colatitudes depending on the size of the planet magnetosphere (\textit{cf} Eq.~\ref{eq.alpha0}, \citealt{2011MNRAS.414.1573V}). The frequencies are 18, 114, and 240 MHz for polar magnetic fields of 10, 50, and 100~G, respectively. The polar-cap boundaries extend down to latitudes of $\pm 29^{\rm o}$,  $\pm 48^{\rm o}$ , and $\pm 53^{\rm o}$, respectively. The magnetic field intensity at these latitudes is shown in the right axis of Figure \ref{fig.rm}b. For comparison, at the Earth, the size of the polar cap is about $17$ to $20^{\rm o}$ \citep{2009AnGeo..27.2913M}, that is, extending to latitudes of $\pm 70$ to $\pm 73^{\rm o}$, showing once more the  extreme case of the magnetosphere of \vtau\ b.

\section{Stellar versus exoplanetary radio emissions}\label{sec.radiowind}
If \vtau\ b is indeed a magnetised planet, it can emit at radio wavelengths due to a physical process called electron cyclotron maser instability (Section \ref{sec.radiomodel}).  However, \vtau\ b may not be the only radio emitter in this exoplanetary system. The hot plasma from the stellar wind can also emit at radio wavelengths, although through a different physical process. Here, we estimate radio emission from the host star's wind.

\subsection{Radio emission from the stellar wind}
A thermal ionised plasma emits bremsstrahlung (free-free) radiation across the electromagnetic spectrum. In the case of stellar winds, free-free radio emission is more intense in the innermost regions of the stellar wind, where the densities are higher \citep{1975A&A....39....1P,1975MNRAS.170...41W,1996ApJ...462L..91L, 2002ARA&A..40..217G}. For large enough densities, these regions can become optically thick to radio wavelengths and, if a planet is embedded in this region, then it is possible that most of the planetary radio emission gets absorbed and does not escape. 

A simple way to calculate  the radio emission of stellar winds was presented in \citet{1975A&A....39....1P}, who assume that the wind is spherically symmetric and isothermal with temperature $T$. These hypotheses are, however, not true in our simulations: the asymmetric distribution of surface magnetic fields generates an asymmetric stellar wind \citep{2014MNRAS.438.1162V}. We also adopt a polytropic wind model, such that the wind temperature is not constant. To take into account the asymmetries of the stellar wind, a radiative transfer calculation performed on each cell of the numerical grid and integrated along the line-of-sight would provide more accurate predictions for the radio-emitting wind. This detailed study will be done in a future work. In the present paper, we compute the free-free emission of the inner regions of the wind of \vtau\  using the model developed by \citet{1975A&A....39....1P} coupled to our simulation results. We caution that these computations should be considered as  estimates of the radio emission.

The stellar wind radio flux density at a  frequency $\nu$ is given by \citep{1975A&A....39....1P}
\begin{eqnarray}\label{eq.Snu}
 S_{\nu} = 10^{-29}  A(\alpha) R_\star^2  \left[ 5.624\e{-28 }I(\alpha) n_0^2 R_\star \right]^{\frac{2}{2\alpha -1}} \nonumber \\
  \left( \frac{\nu}{10 \mathrm{GHz}} \right)^{\frac{-4.2}{2\alpha -1}+2}\left( \frac{T}{10^4 \mathrm{K}} \right)^{\frac{-2.7}{2\alpha -1} +1} \left( \frac{d}{1 \textrm{kpc}} \right)^{-2} \mathrm{mJy} , 
 \end{eqnarray}
 where  the functions $I(\alpha)$ and $A(\alpha)$ are given by
\begin{equation}
I(\alpha) = \int_{0}^{\pi/2} (\sin \theta)^{2(\alpha -1)} \mathrm{d} \theta ,
\end{equation}
 \begin{equation}
A(\alpha)=1 + 2 \sum_{j=1}^{\infty} (-1)^{j+1} \frac{\tau_c^j}{j! j (2 \alpha -1) - 2} , 
\end{equation}
 and $\tau_c=3$.  The wind density is assumed to decay as a power-law with exponent $\alpha$
\begin{equation}
n = n_0 \left(\frac{R_\star}{r}\right)^\alpha ,
\end{equation}
with $n_0$ being the wind base density. The distance $ R_{\nu}$ within which half of the emission $S_\nu$ is produced is
\begin{equation}\label{eq.rnu}
 \frac{R_{\nu}}{R_\star} = \left[ 4.23 \times10^{-27} I(\alpha) n_0^2 R_\star \right]^{\frac{1}{2\alpha -1}} \left( \frac{\nu}{10 \mathrm{GHz}} \right)^{\frac{-2.1}{2\alpha -1}}\left( \frac{T}{10^4 \mathrm{K}} \right)^{\frac{-1.35}{2\alpha -1}} .
 \end{equation}

For the wind temperature $T$ in Equations (\ref{eq.Snu}) and (\ref{eq.rnu}), we use the temperature adopted at the wind base ($10^6$~K) and we use the base density $n_0=10^{12}$cm$^{-3}$ adopted in the model described in Section \ref{sec.windmodel}. To estimate $\alpha$,  we compute a power-law fit for regions within  $r<20R_\star$, which resulted in $\alpha\simeq 3.08$.\footnote{Note that, in the case of a stellar wind that has reached asymptotic wind speed, $\alpha=2$. This reduces, for example, to the case presented in \citet{2002ARA&A..40..217G}.} With this value of $\alpha$, the  temperature and frequency dependencies of $R_\nu$ become
\begin{equation} \label{eq.r1}
R_\nu \propto  \nu^{-0.40} T^{-0.26} \, ,
\end{equation}
which is flatter than an emitting wind that has reached asymptotic speed ($R_\nu \propto  \nu^{-0.7} T^{-0.45}$, \citealt{2002ARA&A..40..217G}). 

\begin{figure}[ht!]
\includegraphics[width=\columnwidth]{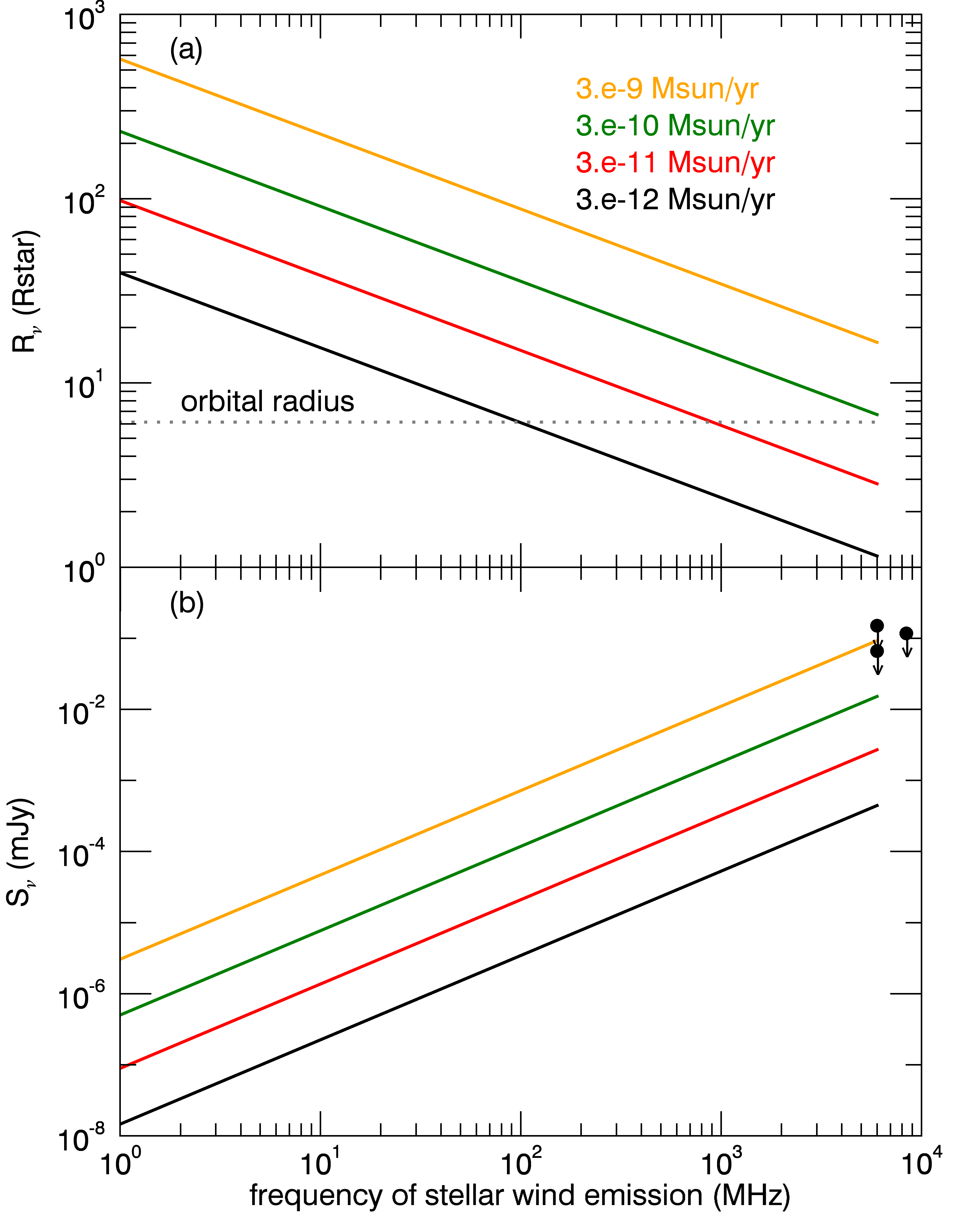}%
\caption{(a) Size of the stellar wind radio-emitting region as a function of frequency for a range of stellar mass-loss rates: $3\e{-12}~\msano$  to  $3\e{-9}~\msano$ from bottom to top solid lines. The dotted line indicates the orbital radius of \vtau\ b.  (b) The same as in (a), but for the flux density of the emitting wind,  $S_\nu$. Circles are the upper limits for the radio emission of \vtau\ b as presented by \citet{bower}.}\label{fig.RSnu}
\end{figure}

Figure \ref{fig.RSnu}a shows the size of the radio-emitting region, where the dotted line  indicates the orbital radius of \vtau\ b. We also explore the effects of the wind mass-loss rates on the radio emission of the wind. For this estimate, we scale the wind densities of the model presented in Section \ref{sec.windmodel}, such that  the mass-loss rates of our parametric study vary between $10^{-12}$ and $3\e{-9}~\msano$. We remind readers that the model presented in Section \ref{sec.windmodel} (our fiducial model) has  $\mdot \sim 3\e{-9}~\msano$. We note that $R_\nu$ gets smaller for higher frequencies and/or lower mass-loss rates. At 10 MHz and for  $\mdot$ in the range $[3\e{-12}, 3\e{-9}]~\msano$, the planet's orbit is always within $R_\nu$, that is, the planet is embedded in the radio-emitting wind. For 100 MHz and in the same range of $\mdot$, $R_\nu \sim 6$ to $90~R_\star$ and $r_{\rm orb}\gtrsim R_\nu$ for $\mdot \lesssim 3\e{-12}~\msano$. For 1000 MHz, $r_{\rm orb}> R_\nu$ for $\mdot \lesssim 3\e{-11}~\msano$.

Regarding the flux density of the emitting wind,  $S_\nu$  increases for higher frequencies and/or higher mass-loss rates. For all the range of $\mdot$ we investigated and the range of frequencies ($[1,6000]$~MHz), the flux density of the emitting wind never reaches the mJy level and $S_\nu \lesssim 0.1$~mJy. Given the low values of $S_\nu$ at low frequencies ($S_\nu \lesssim 10^{-5}$~mJy for $\nu \lesssim 100$~MHz), detecting radio emission from \vtau 's wind at these frequencies with present-day technology is not feasible (Section \ref{sec.conclusion}).

By comparing our radio emission estimates with the observations presented in \citet{bower}, we can place constraints on  the mass-loss rates of \vtau . At certain observing epochs,  \citet{bower} detected \vtau\ in a flaring state, while at other observing epochs, only upper limits for radio emission of \vtau\ could be derived, namely, $<0.066$ and $<0.147$~mJy at 6 GHz (VLA observations) and $<0.117$~mJy at 8.4 GHz (VLBA). These upper limits are shown as circles in Figure \ref{fig.RSnu}b. From that, we infer an upper limit of $\mdot \lesssim 3\e{-9}~\msano$ for the mass-loss rate of \vtau , which is consistent to the value used in our fiducial model (Section \ref{sec.windmodel}). 
 
\subsection{Can radio emission from \vtau\ b propagate through the stellar wind?}
\begin{figure}[ht!]
        \includegraphics[width=\columnwidth]{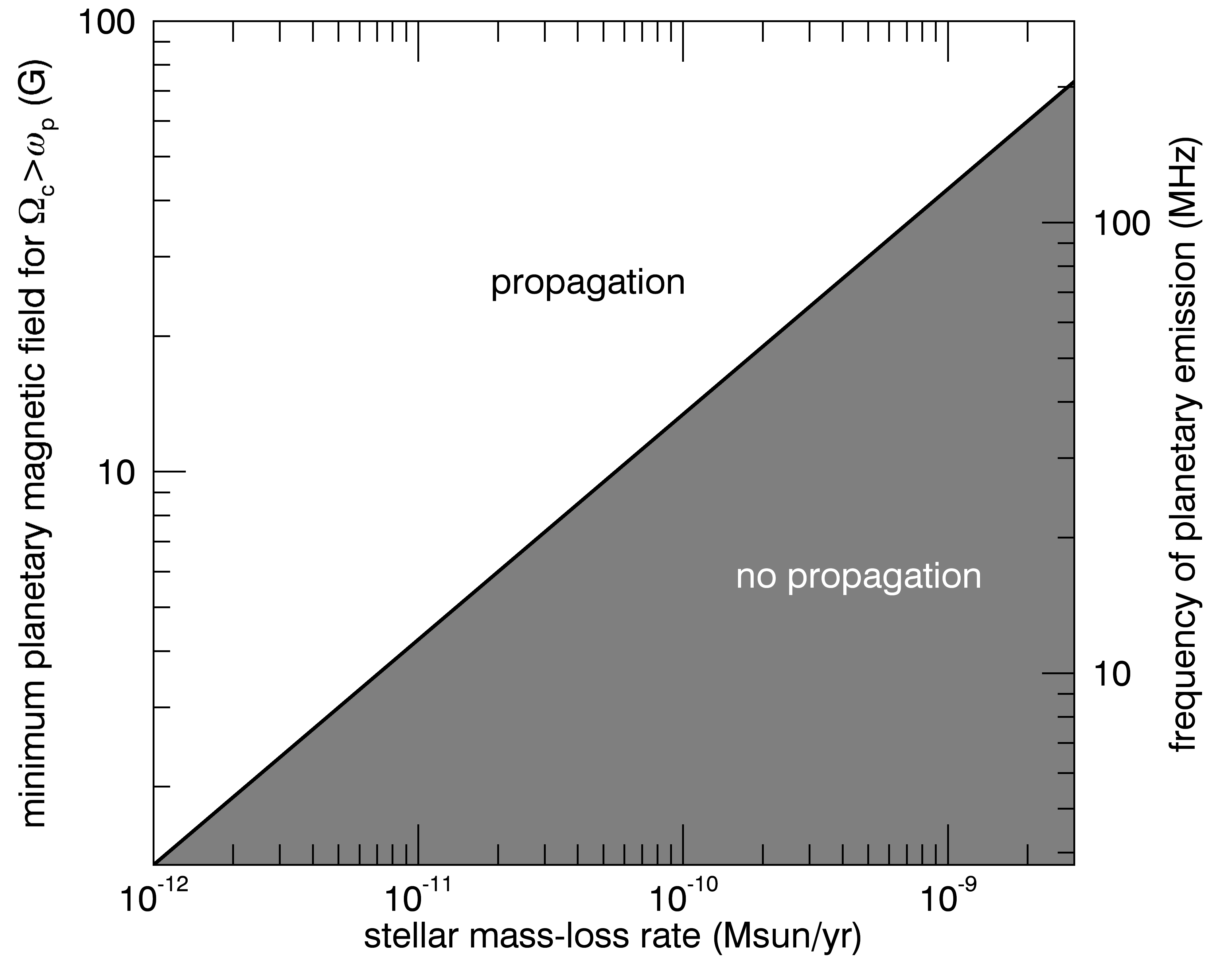}%
\caption{Minimum planetary magnetic field intensity required for the propagation of planetary radio emission through the wind of the host star (Equation \ref{eq.bp}) as a function of stellar wind mass-loss rate. The right hand side axis shows the corresponding frequency of planetary emission (cyclotron).}\label{fig.BpMin}
\end{figure}

The planetary radio emission can only propagate in the stellar wind plasma if the (maximum) frequency of emission $\Omega_c = e B(\alpha_0)/(2 \pi m_e c)$ is larger than the stellar wind plasma frequency  $\omega_p = [n_e e^2/ (\pi m_e)]^{1/2}$ everywhere along the propagation path. In these expressions, $n_e$ is the local electron density of the stellar wind ($n_e=n/2$ for a fully ionised hydrogen wind), $e$  and $m_e$ are the electron charge and mass, respectively, and $B(\alpha_0)$ is the planetary magnetic field at colatitude $\alpha_0$ (half-aperture of the polar-cap boundary). In the scenario in which the planetary emission propagates towards wind lower densities, the condition $\Omega_c > \omega_p$ is met when
\begin{equation}\label{eq.bp}
{B(\alpha_0)} > \left( \frac{n_e}{10^5 ~\textrm{cm}^{-3}} \right)^{1/2} \textrm{G}, 
\end{equation}
where $n_e$ is taken as the electron density at the planetary orbit. Figure \ref{fig.BpMin} shows the minimum planetary magnetic field intensity of \vtau\ b required for the propagation of planetary radio emission through the wind of the host star, as a function of the stellar wind mass-loss rates.

Mass-loss rates of weak-lined T Tauri stars have not been observationally constrained. It is expected that these winds are intermediate between those of classical T Tauri stars with dense winds ($\dot{M}\sim 10^{-10} - 10^{-8}~\msano$, \citealt{1995ApJ...452..736H, 2016RAA....16b..10I}) and zero-age main sequence stars with less dense winds. Very active, main sequence stars can have mass-loss rates of $\dot{M}\sim 2\times 10^{-12}~\msano$ \citep{2005ApJ...628L.143W}. In our work, we estimate an upper limit for the wind of \vtau\ to be $\mdot \lesssim 3\e{-9}~\msano$ (Figure \ref{fig.RSnu}b). Given these values, we speculate that the most probable $\mdot$ for the winds of weak-lined T Tauri stars lies in the range $\sim 10^{-12}$ -- $10^{-10}~\msano$. We use this range of values for the  estimates we present next.\footnote{The non-detection of radio emission from stellar winds can be a useful way to observationally constrain mass-loss rates in solar-type stars, which have rarefied winds \citep[e.g.][]{2000GeoRL..27..501G, 2014ApJ...788..112V,2017arXiv170208393F}.}

For  $\mdot \sim 10^{-12}$ -- $10^{-10}~\msano$, the size of the wind-emitting region ranges between $R_\nu \sim 3$ -- 15 $R_\star$ at $275$~MHz and between $R_\nu \sim 5$ -- 30 $R_\star$ at $50$~MHz (cf.~Figure \ref{fig.RSnu}a). At these frequencies, $R_\nu$ is comparable to the planetary orbital radius ($6.1~ R_\star$), implying that \vtau\ b might be  embedded in the regions of the stellar wind that are optically thick to radio wavelengths, although not so deeply embedded. Given that the radio flux we computed for the planet ($\sim 6$ -- $24$~mJy, Figure \ref{fig.radio}b) is several orders of magnitude larger that the stellar radio flux ($\sim 10^{-6}$ -- $10^{-3}$~mJy, Figure \ref{fig.RSnu}b), it is possible that, even after attenuation, a significant fraction of the planetary radio flux can escape. An accurate radiative transfer calculation of the amount of planetary flux that reaches the observer is left for a more detailed future study. 

For the same range of $\mdot$, planetary radio emission can propagate through the stellar wind if the planetary magnetic field strength is  $\gtrsim1.3$ -- $13$~G (Figure \ref{fig.BpMin}). These minimum values encompass Jupiter's magnetic field intensities and appear to be physically reasonable. The equivalent (minimum) frequency of planetary emission is about $\gtrsim 4$ -- $40$~MHz. At a frequency of $\sim 50$~MHz (see previous paragraph) the wind is (radio) optically thick out to $\sim 5$ to $30~R_\star$ and the planet is not so deeply embedded in the radio-emitting wind. Altogether, these conditions seem very encouraging for planetary radio emission to escape the stellar wind of \vtau\ and be detected at Earth.

\section{Discussion and Conclusions}\label{sec.conclusion}
Based on 3D MHD simulations, we presented here estimates of the radio emission expected for \vtau\ b, the youngest detected exoplanet to date. It orbits an active 2~Myr-old weak-lined T Tauri star at a distance of $0.057$~au, which is about $6.1$ stellar radii. At this distance, the environment surrounding this exoplanet is quite harsh. Using the observationally reconstructed stellar magnetic field, we simulated the wind of this star, allowing us to infer the conditions of the external ambient medium surrounding  \vtau\ b. With these conditions, we then computed the planetary radio emission, using, as analogy, the radiometric Bode's law derived for the magnetised planets in the solar system. According to this empirical law, there exists a relation between the dissipated power of the stellar wind impacting on the magnetosphere of the planet and the power released in the planetary emission. Our model uses simple approximations and should be considered as an initial attempt at calculating the radio flux of \vtau\ b, which is by far currently the best target for detection of exoplanetary radio emission. Detecting \vtau\ b at radio wavelengths would ultimately help us to refine our models.


We showed that, although the frequency of the radio emission is intimately related to the assumed magnetic field of the planet, the radio fluxes are only weakly dependent on that. The estimated  flux densities  from dissipated magnetic wind energy are on the order of $6$ mJy, with peaks at $11$~mJy, for an assumed planetary radius $R_p=R_{\rm jup}$. Given the youth of \vtau\ b, it is likely that this  is a lower limit for the planetary radius. Alternatively, for $R_p=2 R_{\rm jup}$, the radio fluxes we estimated peak at $44$~mJy, with average emission of  $24$~mJy. If \vtau\ b were to have a polar magnetic field intensity of 14.3~G (the maximum value of Jupiter's magnetic field), this means that the emission would occur at a frequency of about 28 MHz, originating at latitudes of about 34$^{\rm o}$, where the magnetic field intensity is $\sim 10 $~G.  Given that LOFAR sensitivity for a 1-h integration time at 20--40 MHz is  3--30 mJy \citep{2011RaSc...46.0F09G}, we conclude that \vtau\ b is an excellent target to look for exoplanetary radio emission with LOFAR. Other present-day instruments that have the potential to detect such radio fluxes in the low-frequency range would be the upgraded UTR-2 \citep{2010A&A...510A..16R} and GMRT \citep{2014A&A...562A.108S}. In the near future, the Square Kilometre Array (SKA)-low array system is expected to outperform LOFAR in terms of sensitivity limits for low frequency ranges  \citep{2015aska.confE.120Z} and it would be an ideal tool for detecting exoplanets at radio wavelengths. 

%

In \citet{2012MNRAS.423.3285V,2015MNRAS.449.4117V},  the radio flux densities of several hot Jupiters (namely: $\tau$ Boo b, HD 46375b, HD 73256b, HD 102195b, HD 130322b, HD 179949b, and HD 189733b) were computed using the same method as here, that is, one in which the data-driven 3D MHD simulations provide the stellar wind characteristics used in the radio flux computations. Compared to the radio flux densities derived in those past works ($0.02$ to $1$ mJy), \vtau\ b presents the best prospect for detecting radio emission (in spite of it not being the closest among the studied systems). This is because \vtau\ is an active star with a large-scale surface magnetic field intensity reaching $700$G \citep{2015MNRAS.453.3706D,2016Natur.534..662D}, up to two orders of magnitude larger than the magnetism of the stars studied in \citet{2015MNRAS.449.4117V}. The latter are main-sequence solar-type stars, which are significantly older than \vtau . In their youth, stars are more active and their magnetism more intense \citep{2014MNRAS.441.2361V}. 

In addition to the variability of radio emission seen along one planetary year (Figs.~\ref{fig.rm} and \ref{fig.radio}, see also \citealt{2010MNRAS.406..409F,2010ApJ...720.1262V, 2012MNRAS.423.3285V,2015MNRAS.449.4117V}), variation within a few years timescale is also expected due to the evolution of the host star magnetism. This means that, within a few years, the conditions of the stellar wind surrounding the planet are likely to change, causing also changes in exoplanetary radio emission. This could be challenging for radio detection,  which will likely require some monitoring of the system.  

The star itself may also contribute at radio frequencies (and potentially through flares that could make the planet modulation harder to detect). We computed thermal radio emission generated in the high density regions of the stellar wind. Comparing our estimated values with the non-detections reported in \citet{bower}, we were able to place a constraint in the mass-loss rate of \vtau : $\mdot \lesssim 3\e{-9}~\msano$. We argued that the most likely values of $\mdot$ lie between $\sim 10^{-12}$ and $10^{-10}~\msano$. With these values, the radio-emitting wind extends to distances of $\sim 3$ to $30~R_\star$ at frequencies of 50 to 275 MHz. This implies that the orbit of \vtau\ b (at $6.1~R_\star$) might sit inside the radio-emitting wind, but does not seem to be deeply embedded. Given that the radio fluxes of the planet are estimated to be $\sim 3$ to 7 orders of magnitude larger than the radio flux of the stellar wind, it is possible that even after undergoing absorption by the stellar wind, a significant fraction of the planetary radio emission can escape the optically (radio) thick wind and reach the observer. Altogether, the \vtau\ system is a very encouraging system for conducting radio observations from both the perspective of radio emission from the planet as well as from the host star's wind.  Equally important, the lack of detection of  emission at radio wavelengths can place important constraints on the characteristics of the yet unobserved winds from weak-lined T Tauri stars.

Finally, exoplanetary radio emission may offer a new avenue for detecting young hot giants. The recent discovery of three of them (\vtau\ b, K2-33b, and TAP26b) may, in fact, indicate that these planets are more frequent than their equivalents around mature Sun-like stars. However, the extreme activity levels of their hosts prevents the use of traditional planet-search techniques. Radio searches can thus boost the findings of young hot Jupiters. It should also give us the chance to probe their magnetic fields from the detected radio frequency, and to simultaneously study stellar winds and star-planet interactions. 


\begin{acknowledgements} 
This work used the BATS-R-US tools developed at the University of Michigan Center for Space Environment Modeling and made available through the NASA Community Coordinated Modeling Center. This work was supported by a grant from the Swiss National Supercomputing Centre (CSCS) under project ID s516. JFD thanks the IDEX initiative of Universit\'e F\'ed\'erale Toulouse Midi-Pyr\'en\'ees for awarding a Chaire d'Attractivit\'e in the framework of which the study of V830 Tau was carried out. The authors would like to thank Manuel Guedel and the anonymous referee for their constructive comments, which greatly improved this manuscript.

 \end{acknowledgements}



\appendix
\section{Detailed derivation of planetary radio flux}\label{sec.apA}

In this work, we assume that radio emission arises near the colatitude $\alpha_0$ of the open-closed magnetic field line boundary; we refer to the region of open field lines as the polar cap.\footnote{Although we refer to $\alpha_0$ as the ``polar cap boundary'',  we note that, due to the harsh conditions imposed by the stellar wind environment, the magnetospheric sizes might be small and, therefore, the polar cap might occupy a large area of the surface of the star (i.e. not only immediately around the poles, as happens for instance with the polar cap area of the Earth).}   We note that the colatitudes of the polar cap  boundary and that of the auroral ring might not match exactly \citep{1975JGR....80.4675S, 2001JGR...106.8101H}, so this is to be considered as a first-order approximation. The half-aperture of the polar cap boundary $\alpha_0$ can be related to the size of the planet magnetosphere as \citep{1975JGR....80.4675S,2010Sci...327.1238T,2011MNRAS.414.1573V,2013ApJ...770...23Z}
\begin{equation}\label{eq.alpha0}
\alpha_0 = \arcsin{[ (R_p/r_M)^{1/2}} ].
\end{equation}
The (dipolar) planetary magnetic field at this colatitude $\alpha_0$ is
\begin{equation}\label{eq.balpha0}
B(\alpha_0) = \frac{B_p}{2} (1+3\cos^2 \alpha_0)^{1/2},
\end{equation}
which corresponds to a maximum electron cyclotron frequency
\begin{equation}\label{eq.fcyc}
f_c =  \frac{e B(\alpha_0) }{2 \pi m_e c},
\end{equation}
where $m_e$ and $e$ are the electron mass and charge, and $c$ is the speed of light (Fig.~\ref{fig.function_variation}). 
Here, we assume that the emission bandwidth $\Delta f$ is approximately the cyclotron frequency \citep{2007P&SS...55..618G}:
\begin{equation}\label{eq.deltafcyc}
\Delta f =  \frac{e B(\alpha_0) }{2 \pi m_e c} = 2.8 \left( \frac{B(\alpha_0)}{1~{\rm G}}\right) ~{\rm MHz} .
\end{equation}

The radio flux is related to the radio power as
\begin{equation}\label{eq.radio}
\phi_{\rm radio} = \frac{P_{\rm radio}}{d^2 \omega \Delta f} \, ,
\end{equation}
where $d$ is the distance to the system and 
\begin{equation}\label{eq.omega}
\omega= 2\times \int_{\alpha_0-\delta\alpha/2}^{\alpha_0+\delta\alpha/2} \sin \alpha \dd \alpha \dd \varphi = 2\times 2 \pi [\cos(\alpha_0-\delta\alpha/2) - \cos (\alpha_0+\delta\alpha/2)]
\end{equation}
is the solid angle of the hollow cone where emission is arising (the factor of two was included in order to account for emission coming from both Northern and Southern hemispheres). The thickness of the hollow cone is assumed to be $\delta\alpha=17.5^{\rm o}$ \citep{2004JGRA..10909S15Z}. 

\subsection{Flux released in the dissipation of the stellar wind kinetic power }
The dissipated kinetic power of the impacting wind on the planet is approximated as the ram pressure of the wind $\rho (\Delta u)^2$ impacting in the magnetospheric cross section $\pi r_M^2$, at a (relative) velocity $\Delta u$ \citep{2007P&SS...55..598Z}
\begin{equation}\label{eq.pwrK}
P_k \simeq \rho (\Delta u)^3 \pi r_M^2 .
\end{equation}
The flux that would be emitted when the kinetic power of the wind gets dissipated in the magnetospheric cross section of the planet - if the radio-magnetic scaling law is effective - can be written as
\begin{equation}
\phi_{\rm radio,kin}  = \frac{P_{\rm radio}}{d^2 \omega \Delta f}  = \frac{\eta_k P_{\rm k}}{d^2 \omega \Delta f} =\frac{\eta_k \rho (\Delta u)^3 \pi r_M^2}{d^2 \omega \Delta f} ,
\end{equation}
where we used Equations~(\ref{eq.radio}) and (\ref{eq.pwrK}). Using Eqs.~(\ref{eq.alpha0}), (\ref{eq.balpha0}), (\ref{eq.deltafcyc}), and (\ref{eq.omega}), the previous expression becomes
\begin{equation}\label{eq.phiradio}
\phi_{\rm radio,kin}  = \frac{\eta_k \rho (\Delta u)^3  R_p^2 [\cos(\alpha_0-\delta\alpha/2) - \cos (\alpha_0+\delta\alpha/2)]^{-1} }{d^2 2 e/[2 \pi m_e c]  {B_p} (1+3\cos^2 \alpha_0)^{1/2}\sin^4{\alpha_0}} \, .
\end{equation}
We note that $\alpha_0$ correlates to $B_p$ through Equations~(\ref{eq.alpha0}) and (\ref{eq.rM}) as
\begin{equation}\label{eq.sinalpha}
B_p = \frac{2 \sqrt{8\pi p_{\rm tot}}}{\sin^6{\alpha_0}} , 
\end{equation}
where $p_{\rm tot}={{(\rho \Delta u^2}  + p+ B^2/8\pi)}$ is the total pressure of the ambient medium external to the planet. Therefore, Eq.~(\ref{eq.phiradio}) becomes
\begin{eqnarray}\label{eq.phiradio2}
\phi_{\rm radio,kin}  &=& \frac{\eta_k \rho (\Delta u)^3 R_p^2 \sin^2{\alpha_0}[\cos(\alpha_0-\delta\alpha/2) - \cos (\alpha_0+\delta\alpha/2)]^{-1}}{d^2 4 e/[2 \pi m_e c ] {\sqrt{8\pi p_{\rm tot}}} (1+3\cos^2 \alpha_0)^{1/2}}  \nonumber \\
  &=& \eta_k \frac{2\pi m_e c}{4 e} \frac{R_p^2}{d^2 }  \frac{\rho (\Delta u)^3}{\sqrt{8\pi p_{\rm tot}}}f(\alpha_0) \, 
\end{eqnarray}
with 
\begin{equation}\label{eq.falpha}
 f(\alpha_0)  =  \frac{\sin^2\alpha_0}{ [\cos(\alpha_0-\delta\alpha/2) - \cos (\alpha_0+\delta\alpha/2)] (1+3\cos^2 \alpha_0)^{1/2}} \, .
\end{equation}

\begin{figure}
\includegraphics[width=\columnwidth]{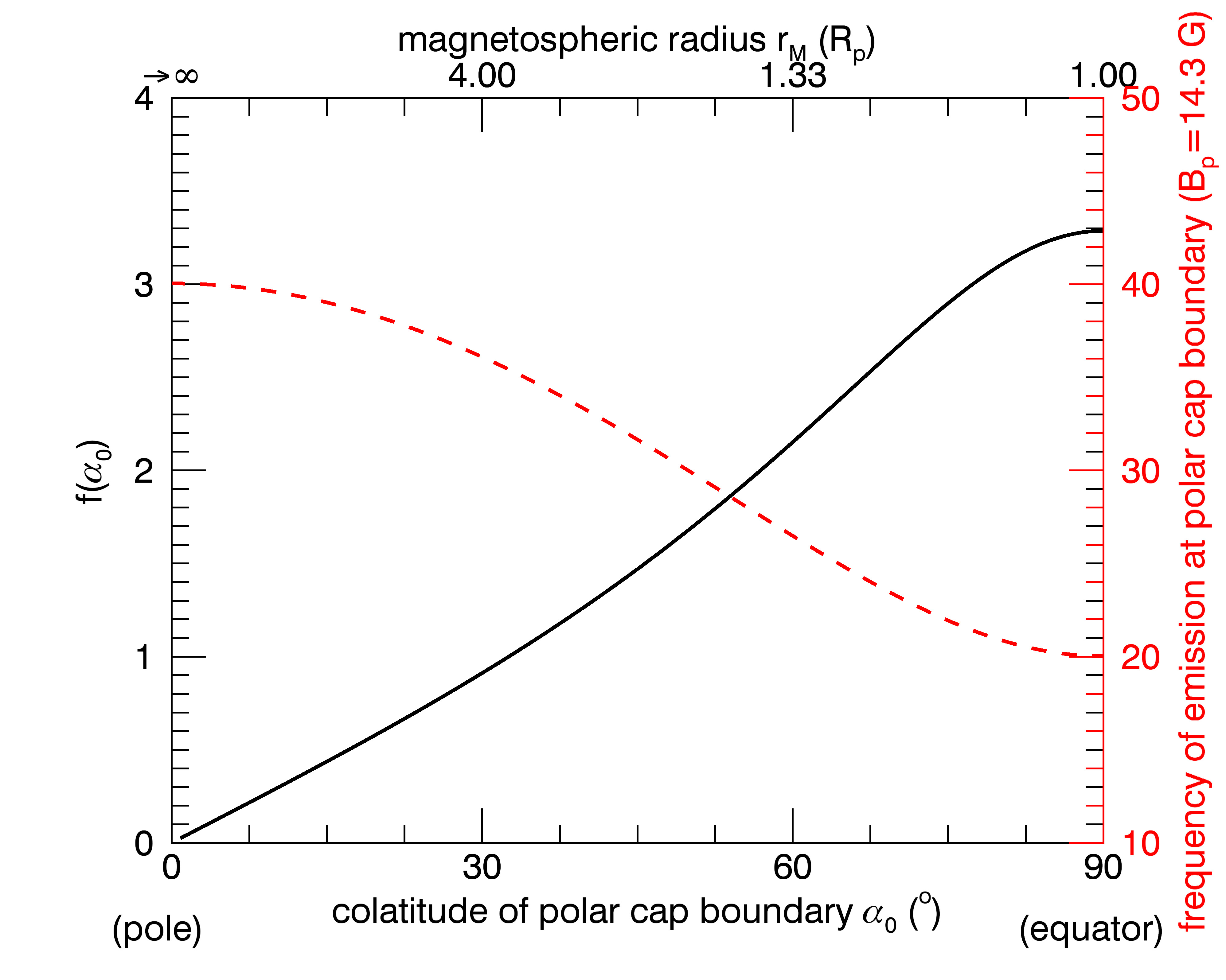}%
\caption{Black solid curve: variation of $f$ (Eq.~\ref{eq.falpha}) as a function of the planetary co-latitude of the polar-cap boundary $\alpha_0$. The top axis indicates the (non-linear) conversion from $\alpha_0$ to a normalised magnetospheric size following Eq.~(\ref{eq.alpha0}). Red dashed curve: the emission frequency at the polar-cap boundary (Eq.~\ref{eq.fcyc}, right axis) for a magnetic field intensity at the pole of 14.3 G. \label{fig.function_variation}}
\end{figure}

We note that the planet magnetic field $B_p$ is no longer explicit in Eq.~(\ref{eq.phiradio2}); the dependence of the $\phi_{\rm radio,kin}$ with $B_p$ is hidden in $\alpha_0$ (Eq.~\ref{eq.sinalpha}) and, consequently, in $f(\alpha_0)$. 
Figure~\ref{fig.function_variation} shows the variation of $f$ as a function of a given co-latitude of the polar-cap boundary $\alpha_0$. As can be seen, this function varies between $0$ and $3.3$ for any aperture angle of the polar-cap boundary. 

If we then group all the constants of Equation~(\ref{eq.phiradio2}) into
\begin{equation}\label{eq.etakprime}
\eta_k^\prime = \eta_k \frac{2\pi m_e c}{4 e \sqrt{8\pi}} \simeq \eta_k 1.8\e{-8} \textrm{\, [cgs~units]}, 
\end{equation}
the radio flux due to the impacting wind (kinetic power) simplifies to
\begin{equation}\label{eq.phiradio3}
\phi_{\rm radio,kin}  = \eta_k^\prime  \frac{R_p^2}{d^2 }\frac{\rho (\Delta u)^3}{{{ p_{\rm tot}}^{1/2}}}  f(\alpha_0),
\end{equation}
which is a function of the angular size of the planet ($R_p/d$), the properties of the ambient medium surrounding the planet (i.e. the stellar wind), and $f(\alpha)$.

\subsection{Flux released in the dissipation of the stellar wind Poynting flux}
The magnetic power $P_B$ can be estimated as the Poynting flux of the stellar wind impacting on the planetary magnetospheric cross-section $S$ \citep{2007P&SS...55..598Z}
\begin{equation}\label{eq.pwrB}
P_B = \int c \frac{{\bf E}\times {\bf B}}{4\pi} \cdot {\rm d}{\bf S} \simeq  \frac{B_{\perp}^2 }{4\pi} (\Delta u) \pi r_M^2\, ,
\end{equation}
where the electric field is ${\bf E} = - \Delta \mathbf{u} \times {\bf B}/c $ and $B_\perp$ is the perpendicular component of the stellar magnetic field to the velocity of the wind particles at the planet's referencial frame. It can be written as
\begin{equation}
B_\perp^2 = B^2 - B_\parallel^2 =  B^2 - \left( \frac{\mathbf{B} \cdot \Delta \mathbf{u}}{|\Delta \mathbf{u}|}\right)^2 .
\end{equation}
The power dissipated in interaction is often ascribed as a fraction $\epsilon$ of Eq.~(\ref{eq.pwrB}), with $\epsilon \sim 0.1$ to $0.2$ \citep{2007P&SS...55..598Z}. However, this efficiency factor is often included in the radiometric Bode's law by means of the $\eta_B$ parameter, such that $P_{\rm radio}=\eta_B P_{\rm B}$, with $\eta_B = 2\e{-3} $ and $P_{\rm B}$ as in Eq.~(\ref{eq.pwrB}).

The flux that is emitted with the dissipation of wind magnetic power can be written as
\begin{equation}
\phi_{\rm radio, mag}  = \frac{P_{\rm radio}}{d^2 \omega \Delta f}  = \frac{\eta_B P_{\rm B}}{d^2 \omega \Delta f}  =  \frac{\eta_B B_{\perp}^2 (\Delta u)  \pi r_M^2}{ 4\pi d^2 \omega \Delta f} ,
\end{equation}
where we used Eqs.~(\ref{eq.pwrB}) and (\ref{eq.radio}). Using Eqs.~(\ref{eq.alpha0}), (\ref{eq.balpha0}), (\ref{eq.deltafcyc}), (\ref{eq.omega}), and (\ref{eq.sinalpha}), the previous expression becomes
\begin{eqnarray}\label{eq.phiradio_B}
\phi_{\rm radio, mag}  &=&  \frac{\eta_B B_\perp^2 (\Delta u)    [\cos(\alpha_0-\delta\alpha/2) - \cos (\alpha_0+\delta\alpha/2)]^{-1} R_p^2 \sin^2{\alpha_0}}{4\pi d^2 4 e/[2\pi m_e c] {\sqrt{8\pi p_{\rm tot}}} (1+3\cos^2 \alpha_0)^{1/2}}  \nonumber \\
  &=& \eta_B \frac{2\pi m_e c}{4 e} \frac{R_p^2}{d^2 }  \frac{B_\perp^2 (\Delta u) }{4\pi \sqrt{8\pi p_{\rm tot}}} f(\alpha_0).
\end{eqnarray}
The magnitude of the planetary magnetic field is again embedded in $f(\alpha_0)$ (\textit{cf} Fig.~\ref{fig.function_variation}). If we then group all the constants of Eq.~(\ref{eq.phiradio_B}) into
\begin{equation}\label{eq.etaBprime}
\eta_B^\prime = \eta_B  \frac{2\pi m_e c}{4 e \sqrt{8\pi}} \frac{1}{4\pi} \simeq \eta_B 1.4\e{-9} \textrm{\, [cgs~units]},
\end{equation}
the radio flux due to the impacting wind (magnetic power) simplifies to
\begin{equation}\label{eq.phiradio_B_2}
\phi_{\rm radio, mag}  = \eta_B^\prime  \frac{R_p^2}{d^2 }\frac{B_\perp^2 (\Delta u)}{{{ p_{\rm tot}}^{1/2}}}  f(\alpha_0),
\end{equation}
which is a function of the angular size of the planet ($R_p/d$), the properties of the ambient medium surrounding the planet (i.e. the stellar wind), and $f(\alpha)$. 

\section{Effects of wind base density in our predicted radio emission}\label{sec.apB}
In our simulations of \vtau\ wind, we noted in Section \ref{sec.windmodel} that the base density  $n_0$ of the wind is a free parameter that affects the computed mass-loss rates and that, with our choice of $n_0$, our computed mass-loss rates were likely near an upper limit of the expected mass-loss rates for weak-lined T Tauri stars. Next, we verify how the radio fluxes estimated in this paper would have changed if the wind densities were to decrease by one to two orders of magnitude, bringing down the wind mass-loss rates of \vtau\ to $\sim 10^{-11}$ -- $10^{-10}~\msano$. 

With the wind densities assumed in this paper, the planet orbit changes from a magnetically-dominated region to a ram-pressure dominated region, as shown in Fig.~\ref{fig.wind}. A decrease in wind density by a factor of $10$ -- $100$ would imply that the planet's orbit would be completely inside the magnetically-dominated region (sub-Alfv\'enic). We remind readers also that, even if a given stellar wind is not dominated by the magnetic pressure, $\phi_{\rm radio, mag}$ can still dominate over $\phi_{\rm radio, kin}$ because of the $3200 \pi^2$ factor in Eq.~(\ref{eq.ratio}). 

To simplify our following estimates, we study two limiting cases: (i) one in which the ram pressure dominates the stellar wind total pressure (similar to the conditions surrounding the Earth) and (ii) one in which the magnetic pressure is the  dominating one (likely to be the condition surrounding \vtau\ b). 
In  case (i), we have that $p_{\rm tot} \to \rho (\Delta u)^2$, and Eqs.~(\ref{eq.phiradio3}) and (\ref{eq.phiradio_B_2}) become
\begin{equation}\label{eq.phiradio4}
\phi_{\rm radio, kin}  \to \eta_k^\prime  \frac{R_p^2}{d^2 }{\rho^{1/2} (\Delta u)^2} f(\alpha_0)\,\,\,\, \textrm{[ram-pressure~dominated]}
\end{equation}
and
\begin{equation}\label{eq.phiradio_B_3}
\phi_{\rm radio, mag}  \to \eta_B^\prime  \frac{R_p^2}{d^2 }\frac{B_\perp^2}{\rho^{1/2}}  f(\alpha_0) \,\,\,\, \textrm{[ram-pressure~dominated]}.
\end{equation}
That is, $\phi_{\rm radio, kin}$ becomes proportional to the square root of the stellar wind density, and $\phi_{\rm radio, mag}$ becomes inversely proportional to $\rho^{1/2}$.  In case (ii), where the magnetic pressure of the stellar wind dominates the total pressure ($p_{\rm tot} \to B^2/(8\pi)$), Eqs.~(\ref{eq.phiradio3}) and (\ref{eq.phiradio_B_2}) become
\begin{equation}\label{eq.phiradio5}
\phi_{\rm radio, kin}  \to \eta_k^\prime  \frac{R_p^2}{d^2 }\sqrt{8\pi}\frac{\rho (\Delta u)^3}{B}  f(\alpha_0)\,\,\,\, \textrm{[magnetic-pressure~dominated]},
\end{equation}
and
\begin{equation}\label{eq.phiradio_B_4}
\phi_{\rm radio, mag}   \to \eta_B^\prime  \frac{R_p^2}{d^2 } \sqrt{8\pi} B \Delta u  f(\alpha_0)\,\,\,\, \textrm{[magnetic-pressure~dominated]} .
\end{equation}
That is, $\phi_{\rm radio, kin}$ becomes linearly proportional to the stellar wind density, while $\phi_{\rm radio, mag}$ becomes independent of that.

Therefore, for planets orbiting farther out (in ram-pressure dominated regions), the stellar wind densities can either increase or decrease radio emission depending on whether it is the kinetic or magnetic flux of the impacting wind that is converted into radio power. On the other hand, for close-in planets orbiting around highly magnetised stars (in the magnetic-pressure dominated region), the radio flux becomes roughly independent of the density, being mainly affected by the local conditions of the stellar magnetic field and the relative velocity of the planet through the stellar wind  (Eq.~\ref{eq.phiradio_B_4}). Since the latter condition is the most likely for \vtau\ b, we conclude that the stellar wind density, an unknown in our numerical simulations, plays a minor role in our estimates of planetary radio fluxes.  

\end{document}